# High-speed ultra-broadband detection based on interfacial work function internal photoemission detector

SIHENG HUANG[1], XIN YUAN[1], XUHONG MA[2], QUAN YU[1], YING LIU[1], CHENJIE PAN[1], CHENG TAN[3], GANGYI XU[3,4], HUA LI[2, *], AND YUEHENG ZHANG[1, *]

[1]*Key Laboratory of Artificial Structures and Quantum Control, School of Physics and Astronomy, Shanghai Jiao Tong University, Shanghai 200240, China*
[2]*Key Laboratory of Terahertz Solid-State Technology, Shanghai Institute of Microsystem and Information Technology, Chinese Academy of Sciences, Shanghai 200050, China*
[3]*Key Laboratory of Infrared Imaging Materials and Detectors, Shanghai Institute of Technical Physics, Chinese Academy of Sciences, Shanghai 200083, China*
[4]*Hangzhou Institute for Advanced Study, University of Chinese Academy of Sciences, Hangzhou 310024, China*
* hua.li@mail.sim.ac.cn, yuehzhang@sjtu.edu.cn

High-speed ultra-broadband detectors play a crucial role in aerospace technology, and national security etc. The interfacial work function internal photoemission (IWIP) detector employs multiple absorption mechanism comprehensively across different wavelength band to achieve complete photon type detection, which makes it possible to realize high-speed and ultra-broadband simultaneously. We propose a ratchet heterojunction IWIP (HEIWIP) detector, which shows 3-165THz ultra-broadband coverage. The high-speed response is investigated in detail by both microwave rectification technology and high-speed modulated terahertz light. Up to 5.1GHz 3dB bandwidth is acquired in terms of microwave rectification measurement. And 4.255GHz inter-mode optical beat note signal was successfully detected.

The rapid development of aerospace technology and national security has posed significant challenges for high-speed detection, as well as multi-band joint detection. High-speed detection plays a pivotal role in heterodyne detection[1], ultrafast spectroscopy[2], 6G sensing-communication integration[3], etc. Meanwhile, multi-band joint detection offers multiple dimensional information compared to single-band detection, leading to improved performance and efficiency. Compact ultra-broadband detectors not only fulfill the demands for multi-band joint detection but also tackle the challenge of balancing between detection requirements and limited system space. With the fast-growing requirement of terahertz (THz) technology, high-speed ultra-broadband detectors that operate from the near infrared to terahertz (THz) range have become a new frontier in detector development.

Currently, the primary high-speed detectors applicable to the terahertz waveband include photo-conductor antenna (PCA[4]), Schottky barrier diodes (SBD[5, 6]), etc., which are predominantly focused on the lower terahertz waveband (<1.6THz). At relatively high terahertz frequencies, terahertz quantum well detector (THz QWP) with a peak frequency of 4.2 THz was reported to detect the 6.2GHz high-frequency modulated terahertz light[7]. However, THz QWPs are constrained to a relatively narrow frequency range, posing certain limitations in terms of detection bandwidth. In addition, due to the restriction of the intersubband transition (ISBT) selection rule for the n-type QWP[8], 45° angle of incidence structure or grid design is typically needed, which is not suitable for large array applications or will complicate device manufacturing. It is worth noting that the THz quantum cascade laser (QCL) can also act as a fast detector for the multiheterodyne dual-comb detection with a bandwidth up to 22GHz[9, 10], nevertheless, it is suitable for narrow band detection either.

In the realm of ultra-broadband detection, the prevailing detector covering the terahertz band is thermal detector, such as the Golay[11] detector. Owing to the influence of the thermal detection mechanism, the response time is limited to milliseconds, which impedes its application in high-speed scenarios[12]. The utilization of two-dimensional materials, such as graphene, hexagonal boron nitride, transition metal dichalcogenides, and even van der Waals materials, offers a promising solution to the slow response speed of conventional thermal detectors[13-15]. Nevertheless, immature materials and fabrication processes imped scalability of production and integration with existing readout circuits. Topological insulators, tellurides, perovskites, and organic materials have also attracted considerable attention for ultra-broadband detection in recent years[16, 17]. However, most of these are based on the thermal effect either, limiting their application for high-speed detection. Additionally, many organic materials are unstable and prone to degradation in air. Therefore, ultra-broadband detectors capable of providing continuous coverage across the near infrared to terahertz range exhibit either slow response time or practical application challenges.

Utilizing multiple absorption mechanism comprehensively across different wavelength with mature semiconductor to achieve entire photon type detection is an effective way to solve this dilemma. Based on such considerations, the homojunction/heterojunction interfacial work function internal photoemission (IWIP)[18, 19] could be well-suited to meet the specified requirements. The involved detection mechanism includes the free carrier absorption (FCA) and inter-valence-band transitions (IVBA) jointly. The photons are absorbed in the emitter layer, then the photo-generated carriers pass through the barrier to the collection layer by internal photoemission. The feature of FCA itself and the integration with IVBA enable ultra-broadband detection from terahertz to near infrared naturally and can absorb vertically incident light without the restriction of selection rule. Recently, a heterojunction interfacial work function internal photoemission (HEIWIP) detector can realize 3-165THz ultra-broadband response at 7K was reported[20]. It should be noted that the detection processes involved in such kind of detector are photon-type rather than thermal-based, which allows for fast response time. The response time of IWIP is estimated to be approximately 20ps, which is close to that of high speed QWIP[21]. At present, there is few research on the high-speed performance of the ultra-broadband IWIP detector. Direct experimental research on the response time is still an open question. Though the current IWIP detector shows high performance for ultra-broadband detection and is potential high speed, the operating temperature of HEIWIP is typically at the level of liquid helium. To enhance the operational temperature, we propose a ratchet HEIWIP by introducing an improved hybrid barrier structure. The ratchet structure effectively suppresses the dark current and achieve temperature of 20K in the terahertz[22].

In this work, the high-speed response of the ultra-broadband ratchet HEIWIP is investigated in detail using both microwave rectification technology and high-speed modulated terahertz light. We have demonstrated 3dB response bandwidth $f_{3dB}$ of the $235 \times 235 \mu m^2$ ratchet HEIWIP detector increases from 1.7GHz to 5.1GHz with increasing the microwave power. A 4.255GHz modulated terahertz light emitted from a terahertz QCL was successfully detected in terms of inter-mode optical beat note spectrum.

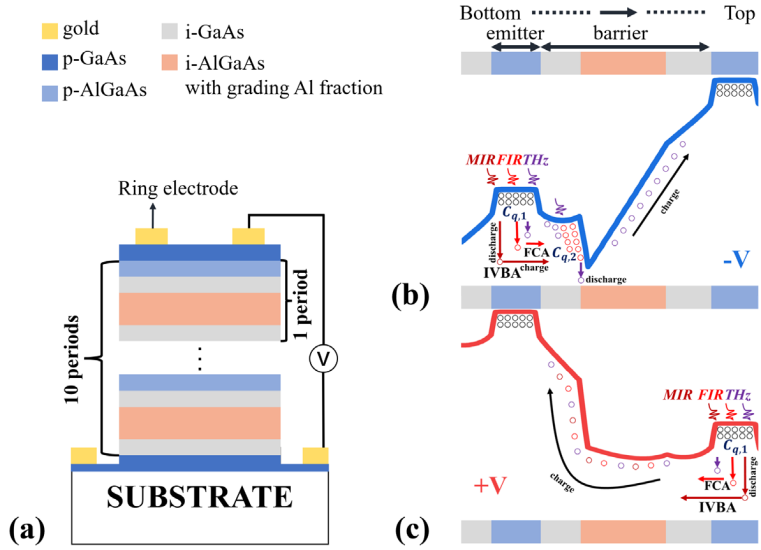

Fig. 1. (a) Device structure of the HEIWIP with ratchets. Band diagram and mechanism of the ratchet under (b) negative bias and (c) positive bias.

The structural diagram of the ratchet HEIWIP detector is shown in Fig. 1(a). The device is grown on semi-insulating GaAs substrates by molecular beam epitaxy (MBE) and processed by wet etching square mesa. From the substrate to the mesa, the function layers are the bottom contact, 10 periods ratchet heterojunction structure, the top contact and the top metal electrode, respectively. Both the bottom and top contacts are p-GaAs and doped with Be to $3.6 \times 10^{18} cm^{-3}$ and covered with Ti\Pt\Au to ensure a good ohmic contact. The period of each ratchet heterojunction structure contains a 500Å p-Al$_{0.005}$Ga$_{0.995}$As emitter layer doped with Be to $3 \times 10^{18} cm^{-3}$, and a hybrid barrier which consists of 1000Å undoped asymmetrical gradient Al$_x$Ga$_{1-x}$As barrier layer sandwiched between two 500Å undoped constant GaAs barriers. The gradient barrier is formed by varying the aluminum fraction of Al$_x$Ga$_{1-x}$As linearly from 0 to 0.07. For comparison, we select three samples A, B and C with respective areas of $235 \times 235 \mu m^2$, $600 \times 600 \mu m^2$ and $1000 \times 1000 \mu m^2$.

The energy band diagrams of the ratchet HEIWIP detector calculated by a Poisson equation solver under negative and positive bias are shown in Fig. 1(b) and (c). Because the ratchet structure in HEIWIP detector breaks the spatial symmetry, the band structure under positive and negative bias is quite different. We define positive bias from mesa to the bottom contact. The ratchet HEIWIP exhibits the similar behavior as the traditional HEIWIP with constant barrier under positive bias. Due to the high concentration of Be doping in the emitter p-Al$_{0.005}$Ga$_{0.995}$As, there are a large number of holes, which absorb the incident THz/IR light and will be excited to a high energy state by both FCA and IVBA. The detector having the advantage of ultra-broadband response that covering THz to IR regions is depend on FCA[23]. The inter-valence-band transitions (IVBA) consists of HH - LH band transitions (LH stands for light hole, HH stands for heavy hole) and LH/HH band – SO band transitions (SO stands for spin-orbit split-off)[20]. The IVBA mainly occurs in the IR region.

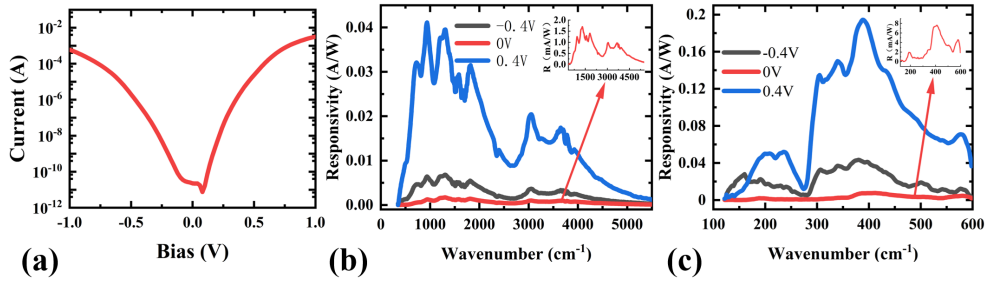

Fig. 2. Device characterizations of the ratchet HEIWIP at 7K. (a) The dark current versus voltage characteristic. (b) IR and (c) THz photoresponse spectra under different biases.

Fig. 2(a) demonstrates the I-V characteristic curves of the sample B at 7K. The I-V characteristic curve exhibits asymmetry due to the inherent asymmetry of the band structure. The ratchets band structure induces directional movement of thermally generated carriers, resulting in producing a net current even at zero bias. IR and THz responsivity spectra of the sample B at 7K under different bias voltages are shown in Fig. 2(b) and (c). The device is an ultra-broadband detector, capable of detecting signals ranging from 3–165THz. Based on the ratchet structure, our THz photoresponse spectra is capable of detecting temperatures up to 20K, while our IR photoresponse spectra can detect temperatures up to 28K and more detail can be examined by referring to our previous research[22].

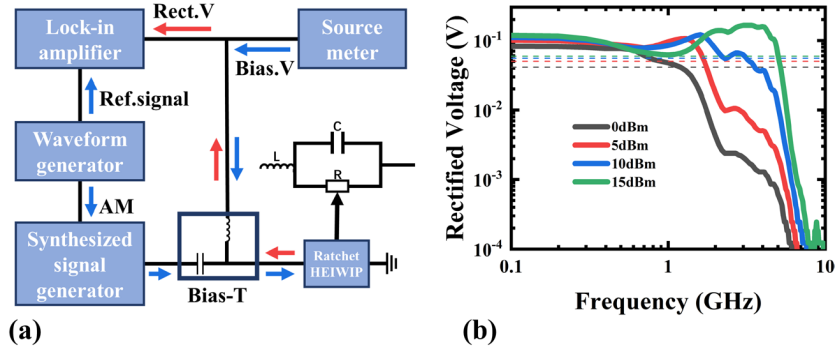

Fig. 3. (a) Schematic of the microwave rectification experiment setup. The inset is the equivalent circuit of the ratchet HEIWIP. (b) Rectified voltage versus frequency for different microwave power as the ratchet HEIWIP is biased at -0.6V. The dashed lines show the 3dB level for different microwave power.

It should be noted that such ratchet HEIWIP is based on photon type detection mechanism completely. Therefore, it is theoretically possible to have a very fast response speed. To evaluate the response time of the ratchet HEIWIP, the microwave rectification measurement is conducted. The fundamental principle is that the time required for the system to return to equilibrium is approximately equal, whether carrier escape occurs due to light excitation or bias perturbation. This time corresponds to the characteristic time that is the carrier lifetime[24]. This characteristic time determines the response speed of the device if we ignore the RC limit. Fig. 3(a) shows the schematic of the microwave rectification experiment setup. The high-frequency microwave signal is provided by the Anritsu series MG3694C synthesized signal generator, which can provide high-frequency modulated signals up to 40GHz. The Agilent 33521A function/arbitrary waveform generator can modulate the high-frequency synthesized signal generator and transmit the modulate information as a reference signal to the Stanford Research Systems model SR830 DSP lock-in amplifier. The DC bias is applied to the device from Keithley 2400 source meter. The K251 wide-band Bias-Tee has three ports, i.e. RF, DC and DC+RF. DC port not only can inject DC voltage to the ratchet HEIWIP but also extract the rectification signal from the device simultaneously. RF port can pass the up to 40GHz modulated high-frequency signal, and at the same time prevent the synthesized signal generator from damaged by prohibiting the reverse input of the DC signal. The DC signal along with the high-frequency modulated signal will be applied to the ratchet HEIWIP through the RF+DC port. In order to minimize the attenuation of the high-frequency modulated signal as much as possible, high performance coaxial RF cable (HUBER + SUHNER SUCOFLEX 102, operating frequency 46GHz) and K connectors are used. The ratchet HEIWIP detector, mounted at the end of a microstrip line, is connected to the line by a wire bond. The microstrip line achieves a 50ohm impedance match by adjusting the thickness of the dielectric layer and the width of the metallic line, and the other end is in contact with the coaxial RF cable. The inset shows the equivalent circuit of the ratchet HEIWIP.

The rectified voltage versus frequency for sample A at different injecting microwave power is shown in Fig. 3(b). The rectified voltage shows a rolling off characteristic of $1/(1+(\omega\tau)^2)$, where $\tau$ is the characteristic time and $\omega$ is the microwave frequency. The measured 3dB frequency is usually used as a measure of the response speed of the device. It can be seen that the 3dB response bandwidth is 1.2GHz at 0dBm and increases from 1.2GHz to 5.1GHz with increasing the microwave power from 0dBm to 15dBm. As evidenced, the response speed is comparable to the narrow band terahertz QWP, whose 3dB bandwidth is 4.3GHz to 5.3GHz in the case of comparable microwave power range[7]. The characteristic time $\tau$ at 0dBm power under -0.4V can be derived as $\tau = 1/(2\pi f_{3dB}) = 94ps$. It is known that the characteristic response time $\tau$ of the device is determined by the slower one of intrinsic carrier lifetime and the RC circuit time.

The carrier lifetime of the ratchet HEIWIP can be roughly estimated[25]. The noise current is related to the mean current through the detector $\bar{I}$[26], $I_{noise}^2 = 4q\bar{I}g_nB$, where $I_{noise}$ is noise current, $g_n$ is the noise gain and B is the bandwidth of the measurement. The photocurrent gain $g_p = \frac{1}{p_cN}$, the ratio $\frac{g_n}{g_p}$ can be approximately equal to $1 - \frac{p_c}{2}$ in the limit $N \gg 1$, N = 10 and $g_n = 1.64$ in our devices. The lifetime $\tau_{life}$ and the carrier transit time across one period $\tau_{trans}$ and the capture probability $p_c$ have a relationship, $p_c = \frac{\tau_{trans}}{\tau_{trans}+\tau_{life}}$. Assuming the effective velocity is $10^7 cm/s$, then $\tau_{transit} = 2.5ps$, so $\tau_{life} = 40ps$, which means the corresponding bandwidth of $f_{life} = 4GHz$.

In order to evaluate the RC time, the device's capacitance at low temperatures and high frequencies is estimated roughly. At low temperatures, carriers hardly penetrate into the barrier, so we can approximately treat the quantum well as metal plates and the barrier act as dielectric layer of a parallel plate capacitor[27]. At high frequency ($\omega \to \infty$ higher than the frequency determined by the inverse relaxation time), the carriers transport process seems to be frozen and the density of states doesn't change[28]. So, it is reasonable to assume the device of capacitance tends to approach the geometrical capacitance, $C_0 = \varepsilon S/d$, where S is the area of the mesa, $\varepsilon$ is the dielectric constant, d is the thickness of the active layers. Therefore, the ratchet HEIWIP detector can be regarded as an equivalent circuit with a capacitance C and resistance R in parallel and an inductance L in series, as shown in the inset of Fig. 3(a). For the small sized sample A, assuming $L \to 0$, the device capacitance C is 2.49pF and the expected RC characteristic frequency is $f_{RC} = 1/(2\pi R_L C) = 1.3GHz$ and $\tau_{RC} = 106ps$.

It can be seen that the carrier lifetime is much shorter than the RC time, indicating that the present ratchet HEIWIP operates under RC mode, which means the response speed can be influenced by the size of detector, as shown by Fig. 4. In order to achieve the response speed determined by the intrinsic carrier lifetime, the size of the detector should be further reduced, nevertheless, the influence of the diffraction limit should be considered carefully. It is worth noting that at higher frequencies than the 3dB point, the rectified voltage still has a high value and can be detected at 10GHz, which means the ratchet HEIWIP detector is potential to have a higher bandwidth.

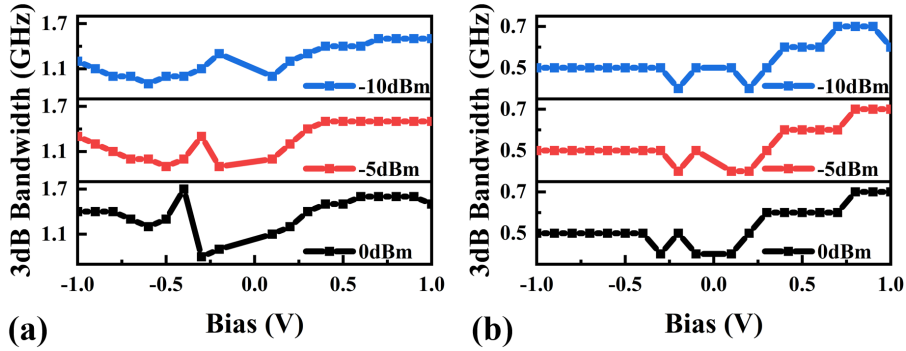

Fig. 4. 3dB bandwidth of the (a) sample A and (b) sample C at different biases for -10dBm, -5dBm, 0dBm power values.

Because different device structure would affect the response speed, the bias voltage dependence of the asymmetric barrier was investigated. The 3dB bandwidth of sample A and sample C at different biases for -10dBm, -5dBm, 0dBm microwave power are demonstrated in Fig. 4(a) and (b). The 3dB bandwidth generally exhibits an increasing trend with the elevation of the electric field under both positive and negative bias conditions, owing to the inverse relationship between capacitance and bias just like QWIP[29]. For the multi-period structure, the distribution of the electric field E along the growth direction is not uniform. Assuming that the average electric field E is given by $V/L$, where V represents the applied bias voltage and L denotes the thickness of the device. At low bias voltage, the electric field near the electrode in the emitter layer is expected to be higher than the average electric field. Conversely, the bulk of the device, the electric field is lower than the average value. Because the majority of the applied voltage drops on the proximity of the electrode area, it is necessary to multiply the structural capacitance $C_0 = \varepsilon S/d$ by a factor of $d/d_e$, where $d_e$ signifies the thickness of the emitter barrier. Therefore, at low bias voltages, the quasistatic capacitance of the ratchet HEIWIP exceeds the geometric capacitance. However, as we increase the bias voltage, both emitter electric field and the bulk of device electric field tend to the average value E, thereby, capacitance gradually decreases and drops to $C_0$ eventually.

It is worth noting that the capacitance of the ratchet HEIWIP detector under positive and negative biases are different thanks to the spatial asymmetry of the energy band. In the absence of excitation, the carriers remain confined within the potential well. However, in cases of thermal or light excitation, the carriers obtain sufficient energy to overcome the interface barrier and be excited to higher energy states. This results in a modification of carrier concentration within the potential well akin to a charge-discharge process. Consequently, apart from geometric capacitance, quantum capacitance also plays a role in devices[30]. The dielectric constant of the host semiconductors dominants the geometrical capacitance and the density of states controls the quantum capacitance[31]. Under negative bias, the ratchet become sharper forming gradient barrier, the carriers that absorb THz or FIR light are impeded by the gradient barrier, leading to an accumulation of carriers in the constant barrier layer. Due to the excitation energy of the carriers in the gradient barrier falling within the terahertz range, irradiation with terahertz light induces an elevation of carrier energy states. Consequently, a charge-discharge structure is established across the gradient barrier, resulting in the formation of extra quantum capacitor $C_{q,2}$ except for $C_{q,1}$. So, there are an additional quantum capacitance under negative bias, resulting in the different capacitance under positive ($C_{q,1}$) and negative bias ($C_{q,1}$ and $C_{q,2}$), as shown in Fig. 1(c) and (b). From Fig. 2(a), it is evident that the IV characteristic curve exhibits asymmetry. The asymmetry of the IV characteristic curve indicates the asymmetry of the capacitance[28], however, the slight asymmetry leads to a minimal difference in capacitance between positive and negative bias.

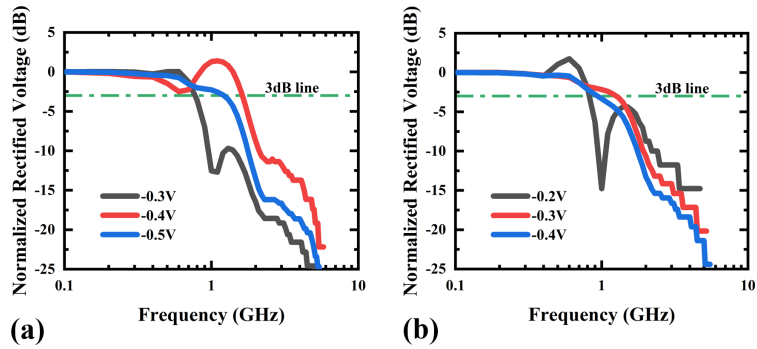

Fig. 5. Normalized rectified voltage versus frequency for sample A at microwave power (a) 0dBm, (b) -5dBm.

In Fig. 4, the resonance peak of the 3dB bandwidth can be observed under a negative bias. For the sample A, the resonance peaks appear at -0.4V, -0.3V, -0.2V for microwave power 0dBm, -5dBm, -10dBm. The bias voltage that results in the resonance peak decreases with the reduction of microwave power. The same scenarios are observed in sample C as well, where the limited bandwidth resulted in less pronounced fluctuations. In order to understand the resonance peak, the normalized rectifications under -0.3V, -0.4V, -0.5V for 0dBm and -0.2V, -0.3V, -0.4V for -5dBm are respectively demonstrated in Fig. 5(a) and (b). For the 0dBm microwave power, when the bias is -0.3V, since the reflection resulted from microstrip line and bonding wire, there is a sharp fall at about 1GHz and a resonance appears at 1.3GHz[32]. When the bias increases to -0.4V, the reflection decreases and the resonance remains, resulting in a maximum 3dB bandwidth. However, as the bias continues to increase, the reflection and resonance disappear, and the rectification curve drops monotonically, just as in the case of positive biases. The rectifications of -5dBm are similar to that of 0dBm, except that due to the reduction of microwave power, the required electric field is also reduced. The rectification curve becomes progressively smoother with decreasing microwave power, and the resonance at -10dBm almost diminishes as expected.

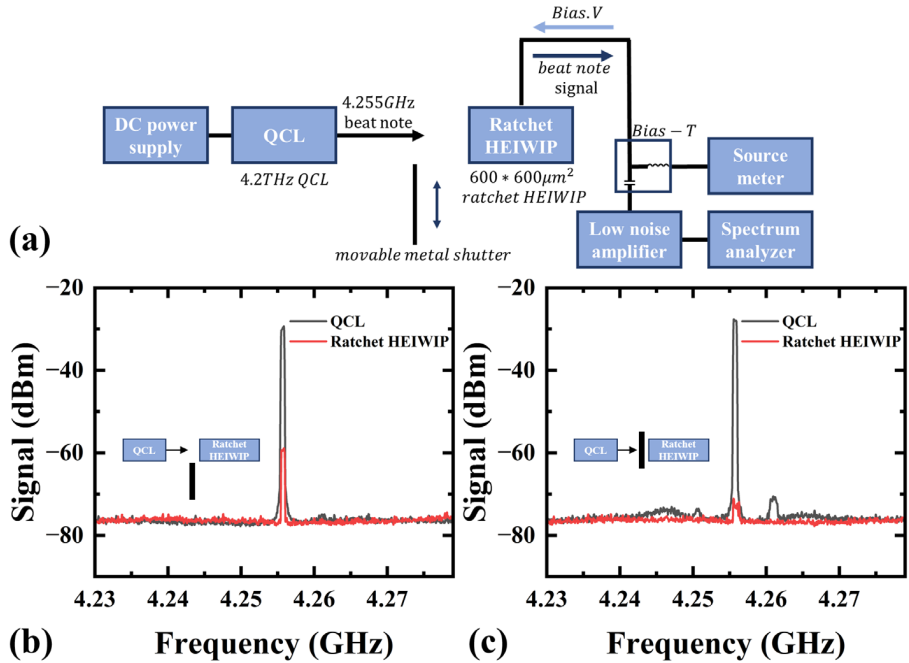

Fig. 6. (a) Schematic experimental setup of the fast terahertz detection. The terahertz quantum cascade laser operating in continuous mode produces modulated terahertz radiation, which is directed vertically onto the ratchet HEIWIP. The optical beat note signal is generated by QCL driven at 930mA. The detector operates at -0.2V, (b) the metal shutter does not block the terahertz light and (c) block the terahertz light.

To further validate the high-speed characteristic directly, a QCL (optical frequency at 4.2THz with beat note frequency at 4.255GHz) is employed as the terahertz source, while the ratchet HEIWIP served as the detector. As ultra-broadband detectors, Fig. 2(c) depicts the frequency of 4.2THz ($140 cm^{-1}$) falling within the detection range of the ratchet HEIWIP. Fig. 6(a) shows the Schematic experimental setup of the fast terahertz detection. Keysight E3644A DC power supply deliver driving currents to the QCL and Keithley 2635B source meter supply bias voltages to the ratchet HEIWIP, respectively. The detected beat note signal is amplified by a low noise amplifier and recorded using the Rohde & Schwarz spectrum analyzer with a bandwidth of 26GHz. A movable metal shutter is employed to enable the ratchet HEIWIP switch to alternate between receiving terahertz light and shield terahertz light. Fig. 6(b) illustrates the RF spectrum measured by the laser and the detector. The beat note signal can be detected by both of the QCL laser itself through self-detecting and the detector. The temperature is set to 13.2K and the resolution bandwidth (RBW) used for this measurement is 100KHz. When the metal shutter is removed away, terahertz light is received by the detector. The optical beat note signal at 4.255GHz with a signal to noise ratio of 47dB for QCL and 17dB for the ratchet HEIWIP are clearly observed, as shown in Fig. 6(b). Due to the linear propagation of optical beat note signals and the omnidirectional propagation of microwave signals in space, a metal shutter is employed to obstruct the terahertz light to demonstrate that the signal detected by the ratchet HEIWIP primarily constitutes an optical beat note signal coupled by the active region rather than a leaking microwave signal coupled by a microstrip line. The result that metal shutter switches to block the terahertz light is shown in Fig. 6(c). When a metal shutter is positioned between the QCL and the detector, the self-detection beat note signal of the QCL experiences slight alteration as the result of the optical feedback effect. The spectral line position remains at 4.255GHz, while the signal strength increases to -27.65dBm, accompanied by a signal to noise ratio of 48dB. Due to complete blockage of terahertz light, only a weak leakage of microwave signals was observed, leading to rapid decay of the beat note signal detected by the ratchet HEIWIP to -71.09dBm with a signal to noise ratio of 5dB. After the terahertz light is shielded by the metal shutter, the QCL beat note signal of self-detection remains essentially unchanged. However, the beat note detected by the ratchet HEIWIP exhibits significant attenuation. This result provides a direct proof that the ratchet HEIWIP can follow the speed at least up to 4.255GHz.

In conclusion, an ultra-broadband and ultra-fast ratchet HEIWIP detector is demonstrated. Utilizing multiple absorption mechanism comprehensively across different wavelength, such detector shows 3-165THz ultra-broadband coverage which spans continuously from terahertz to near infrared. Moreover, it is fully based on photon type detection and capable of detecting terahertz light with a modulation frequency of 4.255GHz. With the implementation of microwave rectification technology, we achieve a 3dB bandwidth ranging from 1.7GHz to 5.1GHz. Compared to other ultra-broadband THz detectors, the ratchet HEIWIP demonstrates an advantage in detection range and response speed, as evidenced by the data presented in Table 1, where the majority of ultra-broadband THz detectors exhibit response times in the microsecond or nanosecond range, whereas the ratchet HEIWIP demonstrates a response speed in the GHz range and the one of the best detectivities. Based on the microwave rectification results of samples A and C, it is evident that the device's response speed is significantly constrained by its area. To further enhance the device's response speed, it is necessary to minimize the device area in order to reduce capacitance and air bridge should be implemented to mitigate parasitic inductance, so as to alleviate the impact of RC limitations.

**Table 1. Comparison of ultra-broadband THz detectors.**

| Device type | Detection range | Detectivity / NEP | Response speed | References |
|---|---|---|---|---|
| 3D graphene FET | UV, VIS, MIR, THz | $2.8*10^{10}\,jones$ | $265ns$ | [33] |
| InBiSe3 | VIS, IR, THz, MMW | $1.35*10^{10}\,jones$ | $10\mu s$ | [34] |
| PtTe2 | 0.02-0.3THz | $47pW/Hz^{\frac{1}{2}}$ | $4.7\mu s$ | [35] |
| HgCdTe | 5-50THz | / | $10MHz$ | [36] |
| Golay cell | 0.04-30THz | $140pW/Hz^{\frac{1}{2}}$ | $50ms$ | [37] |
| ratchet HEIWIP | 3-165THz | $2.5*10^{10}\,jones$ | 4.255GHz | This work |

**Funding.** National Natural Science Foundation of China (12274285, 12074249, 12393833, 62235010, 62325509, 62235019); Ministry of Science and Technology of the People's Republic of China (2023ZD0301000); Shanghai Institute of Microsystem and Information Technology, Chinese Academy of Sciences (ZDBS-LY-JSC009, YSBR-069).

**Disclosures.** The authors declare no conflicts of interest.

**Data availability.** No data were generated or analyzed in the presented research.